\documentclass[12pt]{article}
\usepackage{amssymb, amsmath, amsfonts}

\newtheorem{theorem}{Theorem}

\newtheorem{remark}{Remark}
\newtheorem{example}{Example}

\begin{document}

\begin{center}
{\Large  On the discretization of Darboux Integrable Systems admitting the second-order integrals.}

\bigskip

{Kostyantyn Zheltukhin}\footnote{e-mail: zheltukh@metu.edu.tr}\\

{Department of Mathematics, Middle East Technical University, Ankara, Turkey}\\

\bigskip

{Natalya Zheltukhina}\footnote{e-mail: natalya@fen.bilkent.edu.tr}\\

{Department of Mathematics, Faculty of Science, Bilkent University, Ankara, Turkey}

\end{center}

{\bf Abstract} {\it  We study discretization of Darboux integrable systems. The discretization is done by using $x$- or $y$-integrals of the considered systems.
New examples of semi-discrete Darboux integrable systems are obtained.}


\section{Introduction}

In the present paper we study  the problem of discretization of integrable equations so that the property of integrability is preserved. In particular, we consider hyperbolic type systems
\begin{equation}\label{system_general}
p^i_{xy}=f^i(x,y,p,p_x,p_y)  \qquad i=1,\dots, m\, ,
\end{equation}
where $p=(p^1,\dots, p^m)$, $p_x=(p_x^1,\dots, p_x^m)$ and $p_y=(p_y^1,\dots, p_y^m)$.

For such hyperbolic systems it is convenient to use  Darboux integrability \cite{Dar}. The above system is said to be integrable if it admits $m$ functionally independent  non-trivial $x-$integrals and  $m$ functionally independent non-trivial $y-$integrals. A function $I(x,y,p,p_y, p_{yy}, ...)$ is called an $x-$ integral of the system \eqref{system_general} if
\begin{equation}
D_x I(x,y,p,p_y, p_{yy},...) =0 \qquad \mbox{on all solutions of \eqref{system_general}},
\end{equation}
where $D_x$ is the total derivative with respect to $x$. One can define $y-$ integrals  in a similar way. The Darboux integrable systems are extensively studied, see
\cite{ShYa}-\cite{AFV} and a review paper \cite{ZhMHSh}.

The extension of the notion of Darboux integrability to  discrete and semi-discrete Darboux integrable systems was developed by Habibullin and Pekcan \cite{HP}, see also \cite{Adler}.
In recent years there is an interest in  studying such  systems, see \cite{H}-\cite{ZhZh}.
 A semi-discrete system
 \begin{equation}\label{system_general_discrete}
q^i_{nx}=f^i(x,n,q,q_x,q_1)  \qquad i=1,\dots, m\, ,
\end{equation}
where  $q=(q^1,\dots, q^m)$, $q_x=(q_x^1,\dots, q_x^m)$ and $q_1=(q^1(x,n+1),\dots, q^m(x,n+1))$, is called Darboux integrable if it admits $m$ functionally independent non-trivial $x-$integrals and  $m$ functionally independent non-trivial $n-$integrals.  A function $ J(x,n,q,q_x, q_{xx}, ...)$ is called an $n-$ integral of the system \eqref{system_general_discrete} if
\begin{equation}
D J(x,y,q,q_x, q_{xx}, ...) =J(x,n,q,q_x, q_{xx}, ...) \qquad \mbox{on all solutions of \eqref{system_general_discrete}},
\end{equation}
where $D$ is the shift operator, that is $Dq=q_1$. Note that $Dq_k=q_{k+1}$, $k=1,2,3,...$.
The $x-$integrals  $I(x,n,q,q_1, q_2, ...)$  for the system \eqref{system_general_discrete} are defined in the same way as for continuous system.
 
A hypothesis states that any continuous Darboux integrable system can be discretized with respect to one of the independent variables such that the resulting semi-discrete system
is Darboux integrable and admits the  set of $x-$ or $y-$ integrals of the original system as $n-$integrals \cite{ZhZh}.
The results of our work support the above hypothesis. We complete the discretization of continuous Darboux integrable equations derived by Zhiber and Kostrigina in \cite{ZhK}.
In their paper  Zhiber and Kostrigina  considered the classification problem for continuous Darboux integrable systems admitting
$x-$ and $y-$integrals of the first and second orders. They found all such systems together with  their $x$- and $y$-integrals. Following \cite{ZhK} we have two types of systems.
The first system is
\begin{equation}\label{system42}
\left\{
\begin{array}{l}
u_{xy}=\displaystyle{\frac{u_xu_y}{u+v}+\left(\frac{1}{u+v}+\frac{\alpha }{u+\alpha^2 v}\right)u_xv_y},\\
v_{xy}=\displaystyle{\frac{\alpha^2v_xv_y}{u+\alpha^2v}+\left(\frac{1}{\alpha(u+v)}+\frac{\alpha }{u+\alpha^2 v}\right)u_xv_y}.
\end{array}
\right.
\end{equation}
For $\alpha= 1$ it has $y$-integrals
\begin{equation}\label{case1_Int1}
I=2v -\frac{u+v+c}{u_x}+2c\ln \frac{u_x}{u+v+c},
\end{equation}
\begin{equation}\label{case1_Int2}
J=\frac{u_{xx}}{u_x} - \frac{2u_x+v_x}{(u+v+c)}
\end{equation}
 and the $x$-integrals have the same form in $u,u_y,u_{yy}$ and $v,v_y,v_{yy}$  variables.\\
For $\alpha \neq 1$ it has $y$-integrals
\begin{equation}\label{case1_Int1}
I_1=\left(1+\frac{1}{\alpha}\right)v\left(\frac{u_x}{u+v}\right)^{1-\alpha}-v_x\left(\frac{u_x}{u+v}\right)^{-\alpha},
\end{equation}
\begin{equation}\label{case1_Int2}
J_1=\frac{u_{xx}}{u_x} - \frac{(\alpha+1)u_x+\alpha v_x}{\alpha(u+v)}
\end{equation}
 and the $x$-integrals have the same form in $u,u_y,u_{yy}$ and $v,v_y,v_{yy}$  variables.\\
The second system is
\begin{equation}\label{system43}
\left\{
\begin{array}{l}
u_{xy}=\displaystyle{\frac{vu_xu_y}{uv+d}+\left(\frac{1}{uv+d}+\frac{1 }{\alpha(uv+c)}\right)uu_xv_y},\\
v_{xy}=\displaystyle{\frac{uv_xv_y}{uv+c}+\left(\frac{\alpha}{uv+d}+\frac{1}{uv+c}\right)vu_xv_y}.
\end{array}
\right.
\end{equation}
For $\alpha= 1$ it has $y$-integrals
\begin{equation}\label{case2_Int1}
I_2=\frac{(d-c)v^2u_x^2}{2(uv+d)^2}-\frac{cu_xv_x}{uv+d}
\end{equation}
and
\begin{equation}\label{case2_Int2}
J_2=\frac{u_{xx}}{u_x} + \frac{(d-c)vu_x-cuv_x}{c(uv+d)},
\end{equation}
where $c$ and $d$ are non-zero constants and the $x$-integrals have the same form in $u,u_y,u_{yy}$ and $v,v_y,v_{yy}$  variables.\\
For $\alpha\neq 1$ it has $y$-integrals
\begin{equation}\label{case3_Int1}
I_3=\frac{u_x^\beta v_x}{(uv+d)^\beta}+\frac{\beta v^2 u_x^{\beta+1}}{(uv+d)^{\beta+1}}
\end{equation}
and
\begin{equation}\label{case3_Int2}
J_3=-\frac{u_{xx}}{u_x} + \frac{2vu_x+uv_x}{uv+d},
\end{equation}
where $d$ and $\beta\ne 1$ are non-zero constants,  and the $x$-integrals have the same form in $u,u_y,u_{yy}$ and $v,v_y,v_{yy}$  variables.

To discretize the systems \eqref{system42} and \eqref{system43}  we employ  a method introduced by Habibullin et. all \cite{HZhS}, (see also \cite{HZh}-\cite{ZhZh1}).
In this approach one  takes  $x$- or $y$-integrals of a system and looks for a semi-discrete system admitting such integrals as $n$-integrals. In general one gets a set of semi-discrete systems admitting
these $n$-integrals. For all  sets of $y-$ integrals of systems \eqref{system42} and \eqref{system43}  we obtained corresponding semi-discrete systems. In all cases we were able to choose a semi-discrete system that gives the original system in  the continuum limit.  Also in examples where we can write a semi-discrete system explicitly we have shown that the system is Darboux integrable.

The following theorems are formulated for  a hyperbolic type semi-discrete system
\begin{equation}\label{system}
\left\{
\begin{array}{l}
u_{xn}=f(x,n,u,v,u_x,v_x,u_n,v_n),\\
v_{xn}=g(x,n,u,v,u_x,v_x,u_n,v_n),
\end{array}
\right.
\end{equation}
 where variables $u,v$ depend on a continuous variable $x\in{\mathbb R}$ and a discrete variable $n\in {\mathbb N}$.
Note that  the system  \eqref{system42} in the case $\alpha=1$ was  discretized in \cite{ZhZh}.

\begin{theorem}\label{Th1}
Let $\alpha\ne 1$. A system \eqref{system} admits $n$-integrals \eqref{case1_Int1} and  \eqref{case1_Int2}
if and only if it has the form
\begin{equation}
\begin{array}{l}\label{case1}
u_{1x}=\frac{u_1+v_1}{u+v}{\cal D}_1^{\alpha^{-1}}u_x,\\
v_{1x}=\frac{\alpha+1}{\alpha}\frac{v_1{\cal D}_1^{\alpha^{-1}}-v{\cal D}_1}{u+v}u_x + {\cal D}_1v_x.\\
\end{array}
\end{equation}
The function ${\cal D}_1$ is equal to $1$ or given implicitly by $H(K_1,L_1)=0$, where $H$ is any smooth function and
\begin{equation}\label{case1_K1}
  K_1=\frac{\alpha v_1{\cal D}_1^{\alpha^{-1}}-\alpha v{\cal D}_1^{1+\alpha^{-1}}+(1-{\cal D}_1^{\alpha^{-1}})u_1}{({\cal D}_1^{\alpha^{-1}}-1)^{\alpha+1}},
\end{equation}
\begin{equation}\label{case1_K2}
  L_1=\frac{(u_1-{\cal D}_1^{1+\alpha^{-1}}u)e^{{\cal D}_1^{\alpha^{-1}}}}{{\cal D}_1({\cal D}_1^{\alpha^{-1}}-1)^a}+\frac{(-1)^\alpha e^{{\cal D}_1^{\alpha^{-1}}}(\alpha v_1{\cal D}_1^{\alpha^{-1}}-
  \alpha v{\cal D}_1^{1+\alpha^{-1}}+(1-{\cal D}_1^{\alpha^{-1}})u_1)}{({\cal D}_1^{\alpha^{-1}}-1)^{\alpha+1}}.
\end{equation}
\end{theorem}

Let us construct some examples.
\begin{example}
In the  case ${\cal D}_1=1$  the system \eqref{case1} becomes
\begin{equation}\label{sys_D=1_s1}
\begin{array}{l}
u_{1x}=\dfrac{u_1+v_1}{u+v}u_x,  \\
\\
v_{1x}=\left(1+\dfrac{1}{\alpha}\right)\dfrac{v_1-v}{u+v}u_x  +v_x.
\end{array}
\end{equation}
This system is Darboux  integrable. Indeed,  it has two independent non-trivial $n$- integrals \eqref{case1_Int1}, \eqref{case1_Int2} and two independent non-trivial $x$-integrals
\begin{equation}
\mathcal{F}_1=\frac{v - v_1}{v_1 - v_2} \qquad {\mbox{and}} \qquad
\mathcal{F}_2=\frac{u_1 - u}{(v_1 - v)^{\frac{\alpha}{1 + \alpha}}} -
 \alpha(v_1 - v)^{\frac{1}{1 + \alpha}}.
\end{equation}
The $x$-integrals can be found by considering the $x$-algebra corresponding to the system.
\end{example}

\begin{example}
Considering $K_1=0$ and $\alpha=-1$ we get ${\cal D}_1=\dfrac{u_1+v_1}{u_1+v}$ . Using \eqref{case1} we get the system
\begin{equation}
\begin{array}{l}
u_{1x}=\dfrac{u_1+v}{u+v}u_x,  \\
\\
v_{1x}=\dfrac{u_1+v_1}{u_1+v}v_x.
\end{array}
\end{equation}
This system is Darboux  integrable. Indeed,  it has two independent non-trivial $n$- integrals \eqref{case1_Int1}, \eqref{case1_Int2} and two independent non-trivial $x$-integrals
\begin{equation}
\mathcal{F}_1=\frac{(u_2 + v_1)(v - v_1)}{(u_1 + v_1)(v_1 - v_2)} \qquad {\mbox{and}} \qquad
\mathcal{F}_2=\frac{(u - u_1)(u_1 + v_1)}{(-u_1 + u_2) (u_1 + v)}.
\end{equation}
\end{example}

\begin{example}
Considering $K_1=0$ and $\alpha=\dfrac{-1}{2}$ we get ${\cal D}_1=\dfrac{4u_1+2v_1}{v+\sqrt{v^2+16u_1^2+8u_1v_1}}$. Using \eqref{case1} we get the system
\begin{equation}
\begin{array}{l}\label{E1}
u_{1x}=\dfrac{u_1+v_1}{u+v}\left(\dfrac{v+\sqrt{v^2+16u_1^2+8u_1v_1}}{4u_1+2v_1}\right)^2u_x,  \\
v_{1x}=-\left(\dfrac{v_1}{u+v}\left(\dfrac{v+\sqrt{v^2+16u_1^2+8u_1v_1}}{4u_1+2v_1}\right)^2- \dfrac{v(4u_1+2v_1)}{(u+v)(v+\sqrt{v^2+16u_1^2+8u_1v_1})} \right)u_x\\
\qquad + \dfrac{4u_1+2v_1}{v+\sqrt{v^2+16u_1^2+8u_1v_1}}v_x\\
\end{array}
\end{equation}
This system  has two independent non-trivial $n$- integrals \eqref{case1_Int1} and \eqref{case1_Int2}.
\end{example}

\begin{example}
Considering $L_1=0$ and $\alpha=\dfrac{-1}{2}$ we get ${\cal D}_1=\dfrac{v_1+\sqrt{v_1^2+16u^2+8uv}}{2v+4u}$. Using \eqref{case1} we get the system
\begin{equation}
\begin{array}{l}\label{E2}
u_{1x}=\dfrac{u_1+v_1}{u+v}\left( \dfrac{2v+4u}{v_1+\sqrt{v_1^2+16u^2+8uv}}   \right)^2u_x,  \\
v_{1x}=-\left(\dfrac{v_1}{u+v}\left( \dfrac{2v+4u}{v_1+\sqrt{v_1^2+16u^2+8uv}} \right)^2- \dfrac{v(v_1+\sqrt{v_1^2+16u^2+8uv})}{(u+v)(2v+4u)} \right)u_x\\
\qquad +\dfrac{v_1+\sqrt{v_1^2+16u^2+8uv}}{2v+4u}v_x\\
\end{array}
\end{equation}
This system  has two independent non-trivial $n$- integrals \eqref{case1_Int1} and  \eqref{case1_Int2}.
\end{example}

\begin{remark} In both previous examples  let us consider the corresponding $x$-rings. Denote by $X=D_x$, $Y_1=\displaystyle{\frac{\partial}{\partial u_x}}$,
  $Y_2=\displaystyle{\frac{\partial}{\partial v_x}}$, $E_1=\displaystyle{[Y_1,X]}$, $E_2=\displaystyle{[Y_2,X]}$, $E_3=[E_1, E_2]$.
  Note that $X=u_xE_1+v_xE_2+Y_1+Y_2$. The following multiplication table
  $$\begin{array}{l|c|c|c}
[E_i, E_j]&E_1&E_2&E_3\\
\hline
E_1&0&E_3&-2(u+v)^{-1} E_3\\
\hline
E_2&-E_3&0&-2(u+v)^{-1} E_3\\
\hline
E_3&2(u+v)^{-1} E_3&2(u+v)^{-1} E_3&0
\end{array}
$$
shows that $x$-rings are finite-dimensional. Therefore systems (\ref{E1}) and (\ref{E2}) are Darboux integrable.

\end{remark}


\begin{remark}
Expansion of the function $D_1^{\alpha^{-1}}$ given implicitly by $H(K_1,L_1)=L_1=0$, that is by
$$
\alpha v_1 D_1^{\alpha^{-1}}-\alpha v D_1^{1+\alpha^{-1}}+(1-D_1^{\alpha^{-1}})u_1=0,
$$
into a series of the form $\displaystyle{D_1^{\alpha^{-1}}(u_1, v,v_1)=\sum\limits_{n=0}^\infty a_n(v_1-v)^n}$, where coefficients $a_n$ depend on variables
$u_1$ and $v$ only, yields
$$
\displaystyle{D_1^{\alpha^{-1}}(u_1, v,v_1)=1+\frac{\alpha}{u_1+\alpha^2 v}(v_1-v)+\sum\limits_{n=2}^\infty a_n(v_1-v)^n}
$$
and
$$
\displaystyle{D_1(u_1, v,v_1)= 1+\frac{\alpha^2}{u_1+\alpha^2 v}(v_1-v)+\sum\limits_{n=2}^\infty a_n(v_1-v)^n}.
$$
By letting $u_1=u+\varepsilon u_y$, $v_1=v+\varepsilon v_y$ and taking $\varepsilon \to 0$ one can see that the system   (\ref{case1}) becomes  (\ref{system42}).
\end{remark}

\begin{theorem}\label{Th2}

A system \eqref{system} admits $n$-integrals \eqref{case2_Int1} and \eqref{case2_Int2}
if and only if it has the form
\begin{equation}
\begin{array}{l}\label{case2}
u_{1x}=\dfrac{v(u_1v_1+d){\cal D}_2}{v_1(uv+d)}u_x,\\
\\
v_{1x}=\dfrac{(d-c)vv_1({\cal D}_2^{2}-1)}{2c(uv+d){\cal D}_2}u_x + \dfrac{v_1}{v{\cal D}_2}v_x.\\
\end{array}
\end{equation}
The function ${{\cal{D}}_2}$ is given implicitly by $H(K_2,L_2)=0$, where $H$ is any smooth function and
\begin{equation}\label{case2_K2L2}
  K_2=\frac{v_1({\cal D}_2-1)M^{-\frac{2d}{c+d}}}{v{\cal D}_2}\left( -2cdu_1v_1+uv{\cal D}_2M\right),\qquad
  L_2=\frac{v{\cal D}_2M^{\frac{2d}{c+d}}}{v_1},
 \end{equation}
  where
  $$
  M=2cd +\frac{(c+d)({\cal D}_2-1)u_1v_1}{{\cal D}_2}\, .
  $$

\end{theorem}

\begin{example}
Considering $K_2=0$ we can  get ${\cal D}_2=\dfrac{(2cd+(c+d)uv)u_1v_1}{(2cd+(c+d)u_1v_1)uv}$ . Using \eqref{case2} we get the system
\begin{equation}
\begin{array}{l}
u_{1x}=\dfrac{u_1(u_1v_1+d)(2cd+(c+d)uv)}{u(uv+d)(2cd+(c+d)u_1v_1)}u_x,  \\
\\
v_{1x}=\dfrac{(d-c)}{2c(uv+d)}\left\{ \dfrac{u_1v_1^2(2cd+(c+d)uv)}{u(2cd+(c+d)u_1v_1)}-\dfrac{uv^2(2cd+(c+d)u_1v_1)}{u_1(2cd+(c+d)uv)} \right\}u_x\\
\\
\qquad \qquad +\dfrac{u(2cd+(c+d)u_1v_1)}{u_1(2cd+(c+d)uv)}v_x.
\end{array}
\end{equation}
This system  has two independent non-trivial $n$- integrals \eqref{case2_Int1} and \eqref{case2_Int2}.
One can check that this system also has the following two $n$-integrals
$$
I_2^*=\frac{(2cd+(c+d)uv)u_x}{u(uv+d)},
$$
$$
J_2^*=\frac{(c-d)uv^2u_x}{2c(uv+d)(2cd+(c+d)uv)}+\frac{uv_x}{2cd+(c+d)uv}.
$$
Considering the corresponding $x$-algebra we can also find $x$-integrals given by
\begin{equation}
{\cal{F}}_1=\dfrac{u_1}{u}\left(\dfrac{2cd+(c+d)uv}{2cd+(c+d)u_1v_1}\right)^{\frac{c-d}{c+d}}, \qquad
{\cal{F}}_2=\dfrac{u_1v_1-uv}{u_2v_2-uv}.
\end{equation}
\end{example}
\begin{example}
Considering $K_2=0$ we can  also get ${\cal D}_2=\dfrac{(c+d)u_1v_1}{2cd+(c+d)u_1v_1}$ . Using \eqref{case2} we get the system
\begin{equation}
\begin{array}{l}
u_{1x}=\dfrac{(c+d)u_1v(u_1v_1+d)}{(uv+d)(2cd+(c+d)u_1v_1)}u_x,  \\
\\
v_{1x}=\dfrac{(d-c)v}{2c(uv+d)}\left\{ \dfrac{(c+d)u_1v_1^2}{2cd+(c+d)u_1v_1}-\dfrac{2cd+(c+d)u_1v_1}{(c+d)u_1} \right\}u_x
 +\dfrac{2cd+(c+d)u_1v_1}{(c+d)u_1v}v_x.
\end{array}
\end{equation}
This system  has two independent non-trivial $n$- integrals \eqref{case2_Int1} and \eqref{case2_Int2} and two independent $x$-integrals
$${\cal F}_1=\dfrac{1}{c+d}\left(\dfrac{2cd+(c+d)u_1v_1}{vu_1}\right)^{\frac{c+d}{2d}}+\dfrac{u}{u_1}\left( \dfrac{2cd+(c+d)u_1v_1}{vu_1}\right)^{\frac{c-d}{2d}}
$$
and
$$
{\cal F}_2=\dfrac{v_1u_1^{\frac{d-c}{2d}}(2cd+(c+d)u_1v_1)^{\frac{c+d}{2d}}}{v^{\frac{c+d}{2d}}(2cd+(c+d)(u_1v_1+u_2v_2))}.
$$

\end{example}


\begin{remark}
The expansion of the function $D_2$ given implicitly by $\displaystyle{H(K_2,L_2)=L_2-(2cd)^{2d/(c+d)}=0}$
into a series of the form $\displaystyle{D_2(u_1, v,v_1)=\sum\limits_{n=0}^\infty a_n(v_1-v)^n}$, where coefficients $a_n$ depend on variables
$u_1$ and $v$ only, yields
$
\displaystyle{D_2(u_1, v,v_1)=1+\frac{c}{v(uv+c)}(v_1-v)+\sum\limits_{n=2}^\infty a_n(v_1-v)^n}.
$
By letting $u_1=u+\varepsilon u_y$, $v_1=v+\varepsilon v_y$ and taking $\varepsilon \to 0$ one can see that the system   (\ref{case2}) becomes  (\ref{system43}) with $\alpha=-1$.
\end{remark}
\begin{theorem}\label{Th3}

A system \eqref{system} admits $n$-integrals \eqref{case3_Int1} and \eqref{case3_Int2}
if and only if it has form
\begin{equation}
\begin{array}{l}\label{case3}
u_{1x}=\dfrac{u_1v_1+d_1}{{\cal D}_3(uv+d)}u_x,\\
\\
v_{1x}=\left(\dfrac{-\beta v_1^2}{{\cal D}_3(uv+d)} + \dfrac{\beta v^2 {\cal D}_3^\beta}{uv+ d}\right)u_x+{\cal D}_3^\beta v_x.\\
\end{array}
\end{equation}
The function ${\cal D}_3$ is given implicitly by $H(K_3,L_3)=0$, where $H$ is any smooth function and
\begin{equation}\label{case3_K2}
  K_3=\frac{(v_1-v{\cal D}_3^\beta)^{\beta^{-1}}(d_1u-du_1{\cal D}_3)}{{\cal D}_3},
\end{equation}
\begin{equation}\label{case3_L2}
  L_3=(v_1-v{\cal D}_3^\beta)^{(1-\beta)\beta^{-1}}(d_1{\cal D}_3^{\beta-1}-d_1+(\beta-1)u_1(v_1-v{\cal D}_3^\beta)).
\end{equation}
\end{theorem}

\begin{example}
Considering $K_3=0$ we can  get ${\cal D}_3=\dfrac{v_1^{1/\beta}}{v^{1/\beta}}$ . Using \eqref{case3} we get the system
\begin{equation}
\begin{array}{l}
u_{1x}=\dfrac{(u_1v_1+d_1)v^{1/\beta}}{(uv+d)v_1^{1/\beta}}u_x,  \\
\\
v_{1x}=\left\{ -\dfrac{\beta v_1^2v^{1/\beta}}{v_1^{1/\beta}(uv+d)}+\dfrac{\beta v^2v_1}{v(uv+d)} \right\}u_x
 +\dfrac{v_1}{v}v_x.
\end{array}
\end{equation}
This system  has two independent non-trivial $n$- integrals \eqref{case3_Int1} and \eqref{case3_Int2} and two independent $x$-integrals
$${\cal F}_1=\left(1-\left(\dfrac{v_1}{v}\right)^{\frac{1-\beta}{\beta}}\right)\left(-d_1u+du_1\left(\dfrac{v_1}{v}\right)^{\frac{1}{\beta}}\right)^{\beta-1}
$$
and
$$
{\cal F}_2=\dfrac{{v}^{\frac{1-\beta}{\beta}}-{v_2}^{\frac{1-\beta}{\beta}}}{{v}^{\frac{1-\beta}{\beta}}-{v_1}^{\frac{1-\beta}{\beta}}}.
$$
One can check that this system also has the following two $n$-integrals
$$
I_3^*=\frac{v^{1/\beta}u_x}{uv+d}, \qquad
J_3^*=\frac{v_x}{v}+\frac{\beta v u_x}{uv+d}.
$$

\end{example}

\begin{example}

Considering $K_3=0$ we can  also get ${\cal D}_3=\dfrac{d_1 u}{du_1}$ . Using \eqref{case3} we get the system
\begin{equation}
\begin{array}{l}
u_{1x}=\dfrac{(u_1v_1+d_1)du_1}{(uv+d)d_1 u}u_x,  \\
\\
v_{1x}=\left\{ -\dfrac{\beta  d v_1^2u_1}{d_1 u(uv+d)}+\dfrac{\beta d_1^{\beta}v^2u^\beta}{d^\beta u_1^\beta(uv+d)} \right\}u_x
 +\dfrac{d_1^\beta u^\beta}{d^\beta u_1^\beta}v_x.
\end{array}
\end{equation}
This system  has two independent non-trivial $n$- integrals \eqref{case3_Int1} and \eqref{case3_Int2}
 and two independent $x$-integrals
 $$
 {\cal F}_1=\frac{d_1^\beta u^\beta  v-d^\beta u_1^\beta v_1}{d_2^\beta u_1^\beta  v_1-d_1^\beta u_2^\beta v_2}
 $$
 and
 $${\cal F}_2=\frac{(d_1^\beta u^\beta  v-d^\beta u_1^\beta v_1)( dd_1^\beta u^\beta  u-d_1d^\beta u_1^\beta u+(1-\beta)uu_1)}{dd_1 u u_1}.$$
One can check that this system also has the following two $n$-integrals
$$
I_3^{**}=\frac{du_x}{u(uv+d)}, \qquad
J_3^{**}=\frac{u^\beta v_x}{d^\beta}+\frac{\beta v^2 u^\beta u_x}{d^\beta(uv+d)}.
$$
\end{example}

\begin{example}
Considering $L_3=0$ with $\beta=2$ we   get ${\cal D}_3=\dfrac{d_1+R}{2 u_1v}$, where
$$R=\displaystyle{\sqrt{d_1^2+4u_1v(u_1v_1-d_1)}}.$$  Using \eqref{case3} we get the system
\begin{equation}
\begin{array}{l}
u_{1x}=\dfrac{(u_1v_1+d_1)(d_1-R)}{2(uv+d)(d_1-u_1v_1)}u_x,  \\
\\
v_{1x}=\left\{ \dfrac{v_1^2(R-d_1)}{d_1-u_1v_1}+\dfrac{d_1^2+2u_1v(u_1v_1-d_1)+d_1R}{u_1^2} \right\}\dfrac{u_x}{uv+d}
 +\dfrac{d_1^2+2u_1v(u_1v_1-d_1)+d_1R}{2u_1^2v^2}v_x.
\end{array}
\end{equation}
This system  has two independent non-trivial $n$- integrals \eqref{case3_Int1} and \eqref{case3_Int2}.

\end{example}

\begin{example}
Considering $L_3=0$ with $\beta=1/2$ we   get ${\cal D}_3^{1/2}=\dfrac{2d_1+u_1v_1+R}{2 u_1v}$, where
$$R=\displaystyle{\sqrt{(2d_1+u_1v_1)^2-8d_1u_1v}}.$$  Using \eqref{case3} we get the system
\begin{equation}
\begin{array}{l}
u_{1x}=\dfrac{(u_1v_1+d_1)(2d_1+u_1v_1-R)^2}{16d_1^2(uv+d)}u_x,  \\
\\
v_{1x}=\left\{ \dfrac{-v_1^2(2d_1+u_1v_1-R)^2}{32d_1^2}+\dfrac{v(2d_1+u_1v_1+R)}{4u_1} \right\}\dfrac{u_x}{uv+d}
 +\dfrac{2d_1+u_1v_1+R}{2u_1v}v_x.
\end{array}
\end{equation}
This system  has two independent non-trivial $n$- integrals \eqref{case3_Int1} and \eqref{case3_Int2}.

\end{example}

\begin{remark} In both previous examples  the corresponding $x$-rings have the following multiplication table
  $$\begin{array}{l|c|c|c}
[E_i, E_j]&E_1&E_2&E_3\\
\hline
E_1&0&E_3&\frac{-2v}{d+uv} E_3\\
\hline
E_2&-E_3&0&\frac{-2u}{d+uv} E_3\\
\hline
E_3&\frac{2v}{d+uv} E_3&\frac{2u}{d+uv} E_3&0
\end{array}
$$
where fields $X$, $Y_1$, $Y_2$, $E_1$, $E_2$ and $E_3$ are introduced in the same way as in Remark 1. It shows that the $x$-rings are finite-dimensional and the corresponding systems  are Darboux integrable.
\end{remark}


\begin{remark}
The expansion of the function $D_3$ given implicitly by $H(K_3,L_3)=L_3=0$
into a series of the form $\displaystyle{D_3(u_1, v,v_1)=\sum\limits_{n=0}^\infty a_n(v_1-v)^n}$, where coefficients $a_n$ depend on variables
$u_1$ and $v$ only, yields
$$
\displaystyle{D_3(u_1, v,v_1)=1+\frac{u_1}{\beta u_1 v-d_1}(v_1-v)+\sum\limits_{n=2}^\infty a_n(v_1-v)^n}.
$$
By letting $u_1=u+\varepsilon u_y$, $v_1=v+\varepsilon v_y$ and taking $\varepsilon \to 0$ one can see that the system   (\ref{case3}) becomes  (\ref{system43}). Note that $\beta=-\alpha$.
\end{remark}

\section{Proof of Theorem \ref{Th1}}

It follows from the equality $DJ_1=J_1$ that
\begin{equation}
\frac{u_{1xx}}{u_{1x}} -\left(1+\frac{1}{\alpha}\right) \frac{u_{1x}}{u_1+v_1}-\frac{v_{1x}}{u_1+v_1}=\frac{u_{xx}}{u_x} -  \frac{(\alpha+1)u_x+\alpha v_x}{\alpha(u+v)},
\end{equation}
that is
\begin{equation}\label{Int2_eq}
\begin{array}{l}
\displaystyle{\frac{f_x+f_uu_x+f_vv_x+f_{u_1}f+f_{v_1}g +f_{u_x}u_{xx}+f_{v_x}v_{xx}}{f}- \left(1+\frac{1}{\alpha}\right)\frac{f}{u_1+v_1}-\frac{g}{u_1+v_1}= }\\
\displaystyle{\frac{u_{xx}}{u_x} -  \frac{(\alpha+1)u_x+\alpha v_x}{\alpha(u+v)}.}
\end{array}
\end{equation}
By comparing the coefficients by $v_{xx}$ and $u_{xx}$, we get $f_{v_x}=0$ and $\dfrac{f_{u_x}}{f}=\dfrac{1}{u_x}$. Hence
\begin{equation}\label{case1_new_f}
f(x,n,u,v,u_1,v_1,u_x,v_x)=A(x,n,u,v,u_1,v_1)u_x.
\end{equation}
It follows from $DI_1=I_1$ that
\begin{equation}
\left(1+\frac{1}{\alpha}\right)v_1\left(\frac{Au_x}{u_1+v_1}\right)^{1-\alpha}-g\left(\frac{Au_x}{u_1+v_1}\right)^{-\alpha}=\left(1+\frac{1}{\alpha}\right)v\left(\frac{u_x}{u+v}\right)^{1-\alpha}-v_x\left(\frac{u_x}{u+v}\right)^{-\alpha},
\end{equation}
that is
\begin{equation}\label{case1_new_g}
g=\left(1+\frac{1}{\alpha}\right)\left(\frac{Av_1}{u_1+v_1}-\frac{vA^\alpha}{u+v}\left(\frac{u+v}{u_1+v_1}\right)^{\alpha}\right)u_x+\left(A\frac{u+v}{u_1+v_1}\right)^{\alpha}v_x.
\end{equation}
By substituting the expressions for $f$ and $g$ into \eqref{Int2_eq} and
comparing the coefficients by $v_{x}$, $u_{x}$ and the  free term we get
\begin{eqnarray}
& & \frac{A_x}{A}=0, \\
\nonumber
& & \frac{A_u}{A}+A_{u_1}+\left(1+\frac{1}{\alpha}\right)\left[A_{v_1}\frac{v_1}{u_1+v_1}- \frac{vA^\alpha A_{v_1}}{A(u+v)}\left(\frac{u+v}{u_1+v_1}\right)^{\alpha}-
    \frac{A}{u_1+v_1}\right.\\
\label{A_eqn1_s1}
& & \left.-\frac{Av_1}{(u_1+v_1)^2}+\frac{vA^\alpha}{(u+v)(u_1+v_1)}\left(\frac{u+v}{u_1+v_1}\right)^{\alpha} +\frac{1}{u+v}\right] =0,\\
\label{A_eqn2_s1}
& & \frac{A_v}{A}+\frac{A_{v_1}A^\alpha}{A}\left(\frac{u+v}{u_1+v_1}\right)^{\alpha}-\frac{A^\alpha}{u_1+v_1}\left(\frac{u+v}{u_1+v_1}\right)^{\alpha}+\frac{1}{u+v}=0.
\end{eqnarray}
Let
\begin{equation}\label{case1_A}
 {\cal D}_1=\left(\dfrac{u+v}{u_1+v_1}\right)^{\alpha}A^\alpha.
\end{equation}
In terms of the function ${\cal D}_1$ the equations \eqref{A_eqn1_s1} and \eqref{A_eqn2_s1} become
\begin{eqnarray}
\label{D_eqn1_s1}
& & (u+v){{\cal D}_1}_u +(u_1+v_1){\cal D}_1^{\alpha^{-1}}{{\cal D}_1}_{u_1} +\frac{\alpha+1}{\alpha}(v_1{\cal D}_1^{\alpha^{-1}}-v{\cal D}_1){{\cal D}_1}_{v_1}-{\cal D}_1({\cal D}_1^{\alpha^{-1}}-1)=0,\\
\label{D_eqn2_s1}
& & \frac{{{\cal D}_1}_v}{{\cal D}_1}+{{\cal D}_1}_{v_1}=0.
\end{eqnarray}
The set of solutions of the above system is not empty. For example, ${\cal D}_1=1$ is one singular solution  that leads to the Darboux integrable system (\ref{sys_D=1_s1}).
Let ${\cal D}_1\ne 1$. For function  $W=W(u,v,u_1, v_1, {\cal D}_1)$ equations (\ref{D_eqn1_s1}) and (\ref{D_eqn2_s1}) become
\begin{eqnarray}
 (u+v){W}_u +(u_1+v_1){\cal D}_1^{\alpha^{-1}}{W}_{u_1} +\frac{\alpha+1}{\alpha}(v_1{\cal D}_1^{\alpha^{-1}}-v{\cal D}_1){W}_{v_1}+{\cal D}_1({\cal D}_1^{\alpha^{-1}}-1)W_{{\cal D}_1}=0,\\
 \frac{{W_v}}{{\cal D}_1}+{W}_{v_1}=0.
\end{eqnarray}
After the change of variables $\tilde v=v$, ${\tilde v}_1=v_1-v {\cal D}_1$, $\tilde u=u$, ${ \tilde u}_1=u_1$, $\tilde {\cal D}_1={\cal D}_1$ equations above become
$$\begin{array}{l}
({\tilde u}+{\tilde v})W_{\tilde u}+({\tilde u}_1+{\tilde v}_1+{\tilde v}{\tilde {\cal D}_1}){\tilde {\cal D}_1}^{\alpha^{-1}}W_{{\tilde u}_1}+
((1+\frac{1}{\alpha}){\tilde v}_1{\tilde {\cal D}_1}^{\alpha^{-1}}+\frac{1}{\alpha}{\tilde v}({\tilde {\cal D}_1}^{1+\alpha^{-1}}-{\tilde {\cal D}_1 }))W_{{\tilde v}_1}\\
+({\tilde {\cal D}_1}^{1+\alpha^{-1}}-{\tilde {\cal D}_1 })W_{\tilde {\cal D}_1}=0, \\
W_{\tilde v}=0.
\end{array}$$
We differentiate the first equation  with respect to $\tilde v$, use $W_{\tilde v}=0$,  and get two new equations
\begin{eqnarray}
W_{\tilde u}+{\tilde {\cal D}_1}^{1+\alpha^{-1}}W_{{\tilde u}_1}+\frac{1}{\alpha}({\tilde {\cal D}_1}^{1+\alpha^{-1}}-{\tilde {\cal D}_1 })W_{{\tilde v}_1}=0,\\
\tilde u W_{\tilde u}+ ({\tilde u}_1+{\tilde v}_1){\tilde {\cal D}_1}^{\alpha^{-1}}W_{{\tilde u}_1}+\frac{\alpha+1}{\alpha}{\tilde v}_1{\tilde {\cal D}_1}^{\alpha^{-1}}W_{{\tilde v}_1}+
({\tilde {\cal D}_1}^{1+\alpha^{-1}}-{\tilde {\cal D}_1 })W_{\tilde {\cal D}_1}=0.
\end{eqnarray}
After the change of variables $u_1^*={\tilde u}_1-{\tilde {\cal D}_1 }^{1+\alpha^{-1}}{\tilde u}$, $v_1^*=\alpha{\tilde {\cal D}_1 }^{\alpha^{-1}}{{\tilde v}_1}+(1-{\tilde {\cal D}_1 }^{\alpha^{-1}}){{\tilde u}_1} $,
$u^*=\tilde u$, $v^*=\tilde v$ and ${{\cal D}_1^*}={\tilde {\cal D}_1}$ the last system becomes
$$\begin{array}{l}
W_{u^*}=0,\\
\left(({{\cal D}_1^*}^{\alpha^{-1}}+\alpha^{-1}(1-{{\cal D}_1^*}^{-\alpha^{-1}}))u_1^*+\alpha^{-1}v_1^*{{\cal D}_1^*}^{-\alpha^{-1}}\right)W_{u_1^*}+\frac{\alpha+1}{\alpha}v_1^*{{\cal D}_1^*}^{\alpha^{-1}}W_{v^*}\\
+{{\cal D}_1^*}({{\cal D}_1^*}^{\alpha^{-1}}-1)W_{{{\cal D}_1^*}}=0.
\end{array}
$$
The last equation has a general solution  $H(K_1,L_1)=0$, where $K_1$, $L_1$ (rewritten in  old variables)  are given by \eqref{case1_K1}, \eqref{case1_K2} and $H$ is any smooth function.
Now, using the equalities \eqref{case1_A}, \eqref{case1_new_f} and \eqref{case1_new_g} we obtain the system \eqref{case1}. $\Box$

\section{Proof of Theorem \ref{Th2}}
The equality $DJ_2=J_2$ implies
\begin{multline}\label{case2_Int2_eq}
 \frac{f_x+f_uu_x+f_vv_x+f_{u_1}f+f_{v_1}g+f_{u_x}u_{xx}+f_{v_x}v_{xx}}{f} + \frac{(d_1-c_1)v_1f-c_1u_1g}{c_1(u_1v_1+d_1)}\\=\frac{u_{xx}}{u_x} + \frac{(d-c)vu_x-cuv_x}{c(uv+d)}.
\end{multline}
By comparing the coefficients by $u_{xx}$ and $v_{xx}$ in the above equality we get $f_{v_x}=0$ and $\dfrac{f_{u_x}}{f}=\dfrac{1}{u_x}$. Hence
\begin{equation}\label{case2_new_f}
f=A(x,n,u,v,u_1,v_1)u_x.
\end{equation}
Equality $DI_2=I_2$ implies
\begin{equation}\label{case2_Int1_eq}
 \frac{(d_1-c_1)v_1^2A^2u_x}{2(u_1v_1+d_1)^2}-\frac{cAg}{u_1v_1+d_1}=\frac{(d-c)v^2u_x}{2(uv+d)^2}-\frac{cv_x}{uv+d}.
\end{equation}
It follows from \eqref{case2_Int1_eq} that
\begin{equation}\label{case2_new_g}
g=\left(\frac{(d_1-c_1)v_1^2A}{2c_1(u_1v_1+d_1)}-\frac{(d-c)v_1^2(u_1v_1+d_1)}{2c_1A(uv+d)^2}\right)u_x
 +\frac{c(u_1v_1+d_1)}{c_1A(uv+d)}v_x.
\end{equation}
By substituting the expressions for $f$ and $g$ into \eqref{case2_Int2_eq} and comparing  the coefficients by $u_x$, $v_x$ and free term we get
\begin{eqnarray}\nonumber
   \frac{A_x}{A}=0, \\ \nonumber
  \frac{A_u}{A}+A_{u_1}+\left(\frac{A_{v_1}}{A}-\frac{u_1}{u_1v_1+d_1}\right)\left(\frac{(d_1-c_1)v^2_1A}{2c_1(u_1v_1+d_1)}-\frac{(d-c)v^2(u_1v_1+d_1)}{2c_1A(uv+d)^2}\right) \\
 \label{EE}
 +\frac{(d_1-c_1)v_1A}{c_1(u_1v_1+d_1)}-\frac{(d-c)v}{c(uv+d)}=0,\\ \nonumber
  \frac{A_v}{A}+\frac{c(u_1v_1+d_1)}{c_1A(uv+d)}\left( \frac{A_{v_1}}{A}-\frac{u_1}{u_1v_1+d_1}\right)+\frac{u}{uv+d}=0.
\end{eqnarray}
One can check that
$
A=\dfrac{v(u_1v_1+d)}{v_1(uv+d)}
$
is a particular solution provided $d_1=d$ and $c_1=c$. Now assuming that $A\ne\dfrac{v(u_1v_1+d)}{v_1(uv+d)}$ we introduce new function
\begin{equation}\label{case2_D}
{\cal D}_2=\frac{v_1(uv+d)}{v(u_1v_1+d_1)}A .
\end{equation}
In terms of ${\cal D}_2$ the  system (\ref{EE}) becomes
\begin{eqnarray}\nonumber
& & {{\cal D}_2}_x=0, \\ \nonumber
& & (uv+d){{\cal D}_2}_u+\frac{v(u_1v_1+d_1){\cal D}_2}{v_1}{{\cal D}_2}_{u_1}+\frac{vv_1}{2c_1}\left((d_1-c_1){{\cal D}_2}-(d-c){{\cal D}_2}^{-1}\right){{\cal D}_2}_{v_1} \\ \nonumber
& & -\frac{dv}{c}{{\cal D}_2} +\frac{(d_1+c_1)v}{2c_1}{{\cal D}_2}^2+\frac{v(d-c)}{2c_1}=0,\\ \nonumber
& &c_1v{\cal D}_2{{\cal D}_2}_v+cv_1{{\cal D}_2}_{v_1}+(-c{\cal D}_2+c_1{{\cal D}_2}^2) =0.
\end{eqnarray}
For function $W=W(u,v,u_1, v_1, {\cal D}_2)$ the last two equations  become
\begin{eqnarray}\nonumber
& & (uv+d){W}_u+\frac{v(u_1v_1+d_1)}{v_1}{{\cal D}_2}{W}_{u_1}+\frac{vv_1}{2c_1}\left((d_1-c_1){{\cal D}_2}-(d-c){{\cal D}_2}^{-1}\right)W_{v_1} \\ \nonumber
& & +\left(\frac{dv}{c}{{\cal D}_2} -\frac{(d_1+c_1)v}{2c_1}{{\cal D}_2}^2-\frac{v(d-c)}{2c_1}\right)W_{{\cal D}_2}=0,\\ \nonumber
& &c_1v{\cal D}_2{W}_v+cv_1{W}_{v_1}+(c{\cal D}_2-c_1{{\cal D}_2}^2)W_{{\cal D}_2} =0.
\end{eqnarray}
In new variables  $\tilde{u}=u$, $\tilde{u_1}=u_1$, $\tilde{v}=v(c_1{\cal D}_2-c)$, $\tilde{v_1}=v_1(c_1{\cal D}_2-c) {\cal D}_2^{-1}$, $\tilde{{\cal D}_2}={\cal D}_2$ the last system can be rewritten as
\begin{eqnarray}
\nonumber
\left((c_1\tilde{{\cal D}_2}-c)\tilde{u}\tilde{v}+d(c_1\tilde{{\cal D}_2}-c)^2\right)W_{\tilde{u}}+\frac{\tilde{v}}{\tilde{v_1}}\left(\tilde{u_1}\tilde{v_1}\tilde{{\cal D}_2}(c_1\tilde{{\cal D}_2}-c)
+d_1(c_1\tilde{{\cal D}_2}-c)^2\right)W_{\tilde{u_1}}\\ \label{eee}
+\tilde{v}^2\left(\frac{c_1d \tilde{{\cal D}_2}}{c}-\frac{(d_1+c_1)\tilde{{\cal D}_2}^2}{2}+\frac{c-d}{2}\right)W_{\tilde{v}}+\tilde{v}\tilde{v_1}\left(\frac{(d_1-c_1)\tilde{{\cal D}_2}^2}{2}-\frac{cd_1\tilde{{\cal D}_2}}{c_1}+\frac{c+d}{2}\right)W_{\tilde{v_1}}=0,\\ \nonumber
W_{\tilde{{\cal D}_2}}=0.
\end{eqnarray}
Special solutions of (\ref{eee}) may occur only when $\tilde{{\cal D}_2}=c_1/c$.
We differentiate equation (\ref{eee}) with respect to $\tilde{{\cal D}_2}$ three times and get  the following system of three equations
\begin{eqnarray}\label{e1}
(dc^2-c\tilde{u}\tilde{v})W_{\tilde{u}}+c^2d_1\frac{\tilde{v}}{\tilde{v_1}}W_{\tilde{u_1}}+\frac{(c-d)\tilde{v}^2}{2}W_{\tilde{v}}+\frac{(c+d)\tilde{v}\tilde{v_1}}{2}W_{\tilde{v_1}}=0,
\\
\label{e2}
(c_1\tilde{u}\tilde{v}-2dc_1c)W_{\tilde{u}}-(2d_1c_1c\frac{\tilde{v}}{\tilde{v_1}}+c\tilde{u_1}\tilde{v})W_{\tilde{u_1}}+\frac{c_1d\tilde{v}^2}{c}W_{\tilde{v}}-
\frac{cd_1\tilde{v}\tilde{v_1}}{c_1}W_{\tilde{v_1}}=0,
\\
\label{e3}
dc_1^2W_{\tilde{u}}+\left(\frac{d_1c_1^2\tilde{v}}{\tilde{v_1}}+c_1\tilde{u_1}\tilde{v}\right)W_{\tilde{u_1}}-\frac{(d_1+c_1)\tilde{v}^2}{2}W_{\tilde{v}}+\frac{(d_1-c_1)\tilde{v}\tilde{v_1}}{2}W_{\tilde{v_1}}=0,
\end{eqnarray}
that has no solutions if $c_1\ne c$ or $d_1\ne d$.  In case of $c_1=c$ and $d_1=d$ the system becomes
\begin{eqnarray}\label{ee1}
W_{\tilde{u}}-\frac{\tilde{v}^2(2c^2d+(c-d)\tilde{u_1}\tilde{v_1})}{2c(\tilde{u}\tilde{u_1}\tilde{v}\tilde{v_1}+cd(\tilde{u}\tilde{v}-\tilde{u_1}\tilde{v_1}))}W_{\tilde{v}}-
\frac{\tilde{v}\tilde{v_1}(2c^2d+(c+d)\tilde{u_1}\tilde{v_1})}{2c(\tilde{u}\tilde{u_1}\tilde{v}\tilde{v_1}+cd(\tilde{u}\tilde{v}-\tilde{u_1}\tilde{v_1}))}W_{\tilde{v_1}}=0,
\\
\label{ee2}
W_{\tilde{u_1}}-\frac{\tilde{v}\tilde{v_1}(-2c^2d+(c+d)\tilde{u}\tilde{v})}{2c(\tilde{u}\tilde{u_1}\tilde{v}\tilde{v_1}+cd(\tilde{u}\tilde{v}-\tilde{u_1}\tilde{v_1}))}W_{\tilde{v}}+
\frac{\tilde{v_1}^2(2c^2d+(-c+d)\tilde{u}\tilde{v})}{2c(\tilde{u}\tilde{u_1}\tilde{v}\tilde{v_1}+cd(\tilde{u}\tilde{v}-\tilde{u_1}\tilde{v_1}))}W_{\tilde{v_1}}=0.
\end{eqnarray}
After the change of variables $u_1^*=\tilde{u_1}$, $v_1^*=\tilde{v_1}$, $v^*=\dfrac{\tilde{v}}{\tilde{v_1}}\left(2c^2d+(c+d)\tilde{u_1}\tilde{v_1}\right)^{\frac{2d}{c+d}}$,\\
$u^*=\tilde{u}\tilde{v_1}\left(2c^2d+(c+d)\tilde{u_1}\tilde{v_1}\right)^{\frac{c-d}{c+d}}-2c^2d\tilde{u_1}\tilde{v_1}^2\tilde{v}^{-1}\left(2c^2d+(c+d)\tilde{u_1}\tilde{v_1}\right)^{-\frac{2d}{c+d}}$ equations (\ref{ee1}) and (\ref{ee2}) become $W_{v_1^*}=0$ and $W_{u_1^*}=0$ respectively. We rewrite these first integrals in old variables and get that
the general solution is given implicitly by $H(K_2,L_2)=0$, where $H$ is any smooth function and
$K_2$, $L_2$ are given by \eqref{case2_K2L2}. The form of system (\ref{case2}) follows from (\ref{case2_new_f}), (\ref{case2_new_g}) and (\ref{case2_D}). $\Box$

\section{Proof of Theorem \ref{Th3}}
{\bf Proof.}
Equality $DJ_3=J_3$ implies
\begin{equation}\label{case3_Int2_eq}
-\frac{f_x+f_uu_x+f_vv_x+f_{u_1}f+f_{v_1}g+f_{u_x}u_{xx}+f_{v_x}v_{xx}}{f} +  \frac{2v_1f+u_1g}{u_1v_1+d_1}=-\frac{u_{xx}}{u_x} + \frac{2vu_x+uv_x}{uv+d}.
\end{equation}
By comparing the coefficients by $u_{xx}$ and $v_{xx}$ in the above equality we get $f_{v_x}=0$ and $\dfrac{f_{u_x}}{f}=\dfrac{1}{u_x}$. Hence
\begin{equation}\label{case3_new_f}
f=A(x,n,u,v,u_1,v_1)u_x.
\end{equation}
Equality $DI_3=I_3$ implies
\begin{equation}\label{case3_Int1_eq}
\frac{f^\beta g}{(u_1v_1+d_1)^\beta}+\frac{\beta v_1^2 f^{\beta+1}}{(u_1v_1+d_1)^{\beta+1}}=\frac{u_x^\beta v_x}{(uv+d)^\beta}+\frac{\beta v^2 u_x^{\beta+1}}{(uv+d)^{\beta+1}}.
\end{equation}
Let \begin{eqnarray}\label{case3_D}
{\cal D}_3=\dfrac{u_1v_1+d_1}{A(uv+d)}.\end{eqnarray}
 Then from \eqref{case3_Int1_eq} we get
\begin{equation}\label{case3_new_g}
g=\left(-\frac{\beta v_1^2}{{\cal D}_3(uv+d)}+\frac{\beta v^2{\cal D}_3^\beta}{(uv+d)} \right)u_x+{\cal D}_3^\beta v_x.
\end{equation}
The equality \eqref{case3_Int2_eq} in terms of ${\cal D}_3$ takes the  form
\begin{multline}
\frac{{{\cal D}_3}_x}{{\cal D}_3} + \left( \frac{{{\cal D}_3}_{u}}{{\cal D}_3} + \frac{u_1v_1+d_1}{{{\cal D}_3}^2(uv+d)}{{\cal D}_3}_{u_1}
+ \frac{\beta(v^2{\cal D}_3^\beta-v_1^2{\cal D}_3^{-1})}{{\cal D}_3(uv+d)}{{\cal D}_3}_{v_1} +\frac{v_1}{{\cal D}_3(uv+d)}-\frac{v}{(uv+d)}\right)u_x\\
+\left( \frac{{{\cal D}_3}_v}{{\cal D}_3} +{{\cal D}_3}^{\beta-1}{{\cal D}_3}_{v_1}\right)v_x=0.
\end{multline}
By comparing the coefficients by  $u_x$, $v_x$ and free term we get
\begin{eqnarray}
& & {{\cal D}_3}_x=0, \\ \label{eee2}
& & \frac{uv+d}{{\cal D}_3}{{\cal D}_3}_{u} + \frac{u_1v_1+d_1}{{{\cal D}_3}^2}{{\cal D}_3}_{u_1}+
\frac{\beta v^2{\cal D}_3^\beta-\beta v_1^2{\cal D}_3^{-1}}{{\cal D}_3}{{\cal D}_3}_{v_1} +\frac{v_1}{{\cal D}_3}-v=0,\\ \label{eee3}
& & {{\cal D}_3}_v +{{\cal D}_3}^{\beta}{{\cal D}_3}_{v_1}=0.
\end{eqnarray}
In new variables $\tilde{v_1}=v_1-v{\cal D}_3^\beta$, $\tilde{v}=v$, $\tilde{u}=u$, $\tilde{u_1}=u_1$, $\tilde{{\cal D}_3}={\cal D}_3$  equations (\ref{eee3}) and (\ref{eee2}) can be rewritten for function
$W=W(\tilde{u},\tilde{v}, \tilde{u_1}, \tilde{v_1}, \tilde{{\cal D}_3})$ as follows
$$\begin{array}{l}\nonumber
W_{\tilde{v}}=0, \\ \nonumber
\tilde{{\cal D}_3}(\tilde{u}\tilde{v}+d)W_{\tilde{u}}+(\tilde{u_1}(\tilde{v_1}+\tilde{v}\tilde{{\cal D}_3}^\beta) +d_1)W_{\tilde{u_1}}+
\tilde{{\cal D}_3}(\tilde{v}(\tilde{{\cal D}_3}-\tilde{{\cal D}_3}^{\beta})-\tilde{v_1})W_{\tilde{{\cal D}_3}}-\beta \tilde{v_1}(\tilde{v_1}+\tilde{v}\tilde{{\cal D}_3}^\beta)W_{\tilde{v_1}}=0.
\end{array}
$$
We differentiate the last equation with respect to $\tilde{v}$, use the fact that $W_{\tilde{v}}=0$ and get the new system of equations
\begin{eqnarray} \nonumber
\tilde{u}\tilde{{\cal D}_3}W_{\tilde{u}}+\tilde{u_1}\tilde{{\cal D}_3}^\beta W_{\tilde{u_1}}+
(\tilde{{\cal D}_3}^2-\tilde{{\cal D}_3}^{\beta+1})W_{\tilde{{\cal D}_3}}-\beta \tilde{v_1}\tilde{{\cal D}_3}^\beta W_{\tilde{v_1}}=0,
\\
\nonumber
d\tilde{{\cal D}_3}W_{\tilde{u}}+(\tilde{u_1}\tilde{v_1} +d_1)W_{\tilde{u_1}}
-\tilde{{\cal D}_3}\tilde{v_1}W_{\tilde{{\cal D}_3}}-\beta \tilde{v_1}^2W_{\tilde{v_1}}=0,
\end{eqnarray}
that can be rewritten as
\begin{eqnarray} \nonumber
W_{\tilde{u}}+
\frac{d_1\tilde{{\cal D}_3}-d_1\tilde{{\cal D}_3}^\beta+\tilde{{\cal D}_3}\tilde{u_1}\tilde{v_1}}{d_1\tilde{u}-d\tilde{{\cal D}_3}^\beta \tilde{u_1}+\tilde{u}\tilde{u_1}\tilde{v_1}}W_{\tilde{{\cal D}_3}}
-\frac{\beta d_1\tilde{v_1}\tilde{{\cal D}_3}^{\beta-1}}{d_1\tilde{u}-d\tilde{{\cal D}_3}^\beta \tilde{u_1}+\tilde{u}\tilde{u_1}\tilde{v_1}}W_{\tilde{v_1}}=0,
\\
\nonumber
W_{\tilde{u_1}}-
\frac{\tilde{{\cal D}_3}(d\tilde{{\cal D}_3}-d\tilde{{\cal D}_3}^\beta+\tilde{u}\tilde{v_1})}{d_1\tilde{u}-d\tilde{{\cal D}_3}^\beta \tilde{u_1}+\tilde{u}\tilde{u_1}\tilde{v_1}}W_{\tilde{{\cal D}_3}}
+\frac{\beta \tilde{v_1}(d\tilde{{\cal D}_3}^{\beta}-\tilde{u}\tilde{v_1})}{d_1\tilde{u}-d\tilde{{\cal D}_3}^\beta \tilde{u_1}+\tilde{u}\tilde{u_1}\tilde{v_1}}W_{\tilde{v_1}}=0.
\end{eqnarray}
After the change of variables $u^*=\tilde{u}\tilde{v_1}^{1/\beta}d_1^{1/(1-\beta)}\tilde{{\cal D}_3}^{-1}-dd_1^{\beta/(1-\beta)}\tilde{u_1}\tilde{v_1}^{1/\beta}$, \\
${\cal D}_3^*=\tilde{v_1}^{(1-\beta)/\beta}\tilde{{\cal D}_3}^{\beta-1}-\tilde{v_1}^{(1-\beta)/\beta}+(\beta-1)d_1^{-1}\tilde{u_1}\tilde{v_1}^{1/\beta}$,
$u_1^*=\tilde{u_1}$, $v^*=\tilde{v}$, $v_1^*=\tilde{v_1}$ the last two equations become respectively $W_{v_1^*}=0 $ and $W_{u_1^*}=0$.
We rewrite these first integrals in old variables and get that general solution is given implicitly by $H(K_3,L_3)=0$, where $H$ is any smooth function and
$K_3$, $L_3$ are given by \eqref{case3_K2}, \eqref{case3_L2}.  The form of system (\ref{case3}) follows from (\ref{case3_new_f}), (\ref{case3_new_g}) and (\ref{case3_D}). $\Box$

\end{document}